\pdfoutput=1
\documentclass[twocolumn,english,aps,superscriptaddress,prl]{revtex4-1}

\usepackage{babel}
\usepackage{amsmath}
\usepackage{amssymb}
\usepackage{graphicx}

\newcommand{\appropto}{\mathrel{\vcenter{
  \offinterlineskip\halign{\hfil$##$\cr
    \propto\cr\noalign{\kern2pt}\sim\cr\noalign{\kern-2pt}}}}}

\begin{document}

\title{Tunable quantum criticality in multi-component Rydberg arrays}

\author{Natalia Chepiga}
\affiliation{Kavli Institute of Nanoscience, Delft University of Technology, Lorentzweg 1, 2628 CJ Delft, The Netherlands}

\date{\today}
\begin{abstract}
Arrays of Rydberg atoms have appeared as a remarkably rich playground to study quantum phase transitions in one dimension. One of the biggest puzzles that was brought forward in this context are chiral phase transitions out of density waves. Theoretically predicted chiral transition out of period-four phase is still pending experimental verification mainly due to extremely short interval over which this transition is realized in a single-component Rydberg array.  In this letter we show that multi-component Rydberg arrays with extra experimentally tunable parameters  provide a mechanism to manipulate quantum critical properties without breaking translation symmetry explicitly. 
  We consider an effective blockade model of two component Rydberg atoms. Weak and strong components obey nearest- and next-nearest-neighbor blockades correspondingly. When laser detuning is applied to either of the two components the system is in the period-3 and period-2 phases. But laser detuning applied to both components simultaneously  stabilizes  the period-4 phase partly bounded by the chiral transition. We show that relative ratio of the Rabi frequencies of the two components tunes the properties of the conformal Ashkin-Teller point and allows to manipulate an extent of the chiral transition. The prospects of multi-component Rydberg arrays in the context of critical fusion is briefly discussed.
\end{abstract}

\maketitle

%%%%%%%%%%%%%%%%%%%%%%%%%%%%%%%%%%%% INTRODUCTION %%%%%%%%%%%%%%%%%%%%%%%%%%%%%%%%%%%%

% \section{Introduction}
%
{\bf Introduction.} Understanding the nature of quantum phase transitions in low-dimensional systems is a central topic of condensed matter physics\cite{giamarchi,tsvelik}. 
  One of the most debated and long-standing problem in the theory of phase transitions is a chiral melting of the density-wave order that roots back to the study of adsorbed monolayers\cite{Den_Nijs,SelkeExperiment,birgeneau,Selke1982,HuseFisher,HUSE1983363,HuseFisher1984}. Recent experiments on Rydberg atoms attracts new attention to this problem now in the context of one-dimensional (1D) quantum chains\cite{kibble_zureck,scipost_chepiga,prl_chepiga,PhysRevResearch.4.043102,giudici,chepiga2021kibble,PhysRevB.98.205118,samajdar,rader2019floating}.
 The microscopic Hamiltonian of a Rydberg array can be formulated in terms of hard-core bosons:
\begin{eqnarray}
  H_{\text{Ryd}}=\sum_i \left[-\Omega(d^\dagger_i+d_i)-\Delta n_i+\sum_{R=1}^{+\infty} V_R n_in_{i+R}\right]
  \label{eq:hamilt1}
\end{eqnarray}
where $\Omega$ is a Rabi frequency and $V_R\propto R^{-6}$ is the van der Waals potential. 
The competition between the laser detuning $\Delta$ that keeps atom in a Rydberg state and strong van der Waals interaction that blocks simultaneous excitation of multiple atoms within a certain radius realizes a sequence of density-wave lobes with integer periodicities $p=2,3,4,...$ \cite{kibble_zureck,rader2019floating}. These ordered phases are surrounded by the disordered phase with incommensurate short-range order. Incommensurability signals chiral perturbations in a system that have a drastic effect on the nature of quantum phase transitions\cite{HuseFisher,Ostlund,HuseFisher1984}.

For $p=2$ chiral perturbations do not appear and the transition, if continuous, is in the Ising universality class. For $p\geq5$ the transition takes place via an incommensurate Luttinger liquid phase (also known as a floating phase)\cite{rader2019floating}. 
The transition out of $p=3$ phase is more complicated. Inside the disordered phase chiral perturbations vanish along commensurate lines with wave-vectors $q=2\pi/p$. At the point where a commensurate line hits the boundary of the period-3 phase the transition is conformal in the three-state Potts universality class.  Away from this point chiral perturbations are relevant and the transition is believed to be in the Huse-Fisher chiral universality class\cite{HuseFisher1984,samajdar,prl_chepiga,giudici,PhysRevResearch.4.043102}. When chiral perturbations are strong a chiral transition is eventually replaced by a floating phase\cite{prl_chepiga,rader2019floating,PhysRevResearch.4.043102}.

\begin{figure}[h!]
\centering 
\includegraphics[width=0.5\textwidth]{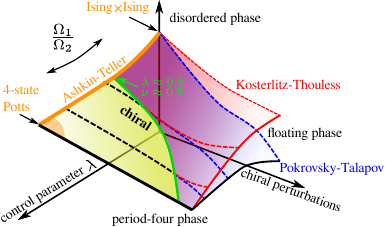}
\caption{{ Nature of the quantum phase transition between the period-four and disordered phases.} Vertical axis states for some relevant operator that brings a system from the disoreder to the period-4 phase. Orange line states for the conformal Ashkin-Teller transition. Yellow and orange regions at finite chiral perturbations correspond to direct chiral and Ashkin-Teller transitions. Blue and red surfaces indicate Pokrovsky-Talapov and Kosterlitz-Thouless transitions with a floating phase between the two. Green line states for the Lifshitz line.}
\label{fig:sketch3d}
\end{figure}

The most intriguing case is $p=4$. Along the commensurate line $q=\pi/2$ the transition is  conformal\cite{chepiga2021kibble} however the underlying Ashkin-Teller critical theory forms a {\it weak} universality class\cite{kohmoto,difrancesco}. This implies that critical exponents, for instance $\nu$ that describes a divergence of the correlation length, can be continuously tuned by an external parameter $\lambda$\cite{kohmoto}. The Ashkin-Teller model  interpolates between two decoupled Ising chains at $\lambda=0$ and the symmetric four-state Potts point at $\lambda=1$ (see Fig.\ref{fig:sketch3d}). The exponent $\nu$ as a function of $\lambda$ is known exactly\cite{kohmoto,obrien}: $$\nu=\frac{1}{2-\frac{\pi}{2}[\arccos(-\lambda)]^{-1}}.$$
As sketched in Fig.\ref{fig:sketch3d} the effect of chiral perturbations that appear in Rydberg arrays away from the commensurate line change the nature of the transition\cite{HuseFisher1984,Den_Nijs,nyckees2022commensurate,luescher}. 
When $\nu\gtrsim 0.8$ chiral perturbations immediately open a floating phase; when $0.8\gtrsim\nu\geq(1+\sqrt{3})/4$  chiral perturbations are also relevant but for a while the transition is direct in the chiral universality class\cite{nyckees2022commensurate,luescher}; finally, when $(1+\sqrt{3})/4>\nu\geq 2/3$ chiral perturbations are irrelevant and for a certain interval the transition is in the Ashkin-Teller universality class that upon increasing chiral perturbations is followed first by the chiral transition and then by the floating phase\cite{schultz,luescher}. In the single-component Rydberg array the conformal Ashkin-Teller point has critical exponent $\nu\approx0.78$ and the second scenario is realized\cite{chepiga2021kibble}. However, as illustrated in Fig.\ref{fig:sketch3d}, the closer is the Ashkin-Teller point to $\nu\approx0.8$ the shorter is the interval of the chiral transition. As a consequence, the chiral transition in a Rydberg array appears only very close to the conformal point\cite{PhysRevResearch.4.043102} making its experimental investigation extremely difficult.

In this letter we directly address this problem and show how  to manipulate quantum critical properties and to enlarge an extent of the chiral transition with two-component bosonic Rydberg array (see below). We show that the ratio between Rabi frequencies of the two components tunes the properties of the Ashkin-Teller multicritical point that in turn controls an appearance and an extent of the chiral transition.

\begin{figure}[t!]
\centering 
\includegraphics[width=0.4\textwidth]{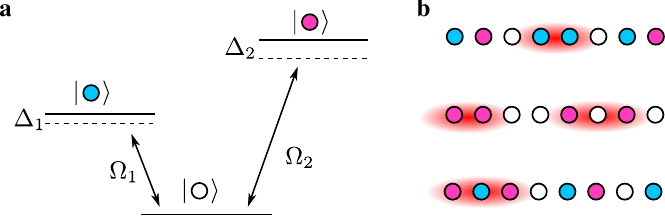}
\caption{Sketch of the two-component Rydberg atom. (a) Atoms are excited to the two Rydberg levels $\alpha=1,2$ from the ground state by a laser with Rabi frequency $\Omega_\alpha$ and detuning $\Delta_\alpha$. (b) Interaction within the strong (magenta) and weak (blue) components result in a nearest- and next-nearest-neighbor blockades.  Configurations that violate these blockades  are marked with red ellipses. }
\label{fig:sketch}
\end{figure}

{\bf Multi-component Rydberg atoms.} Rydberg arrays defined in  Eq.\ref{eq:hamilt1} have two independent parameters while the phase diagram of the chiral Ashkin-Teller model\cite{luescher} sketched in Fig.\ref{fig:sketch3d} requires at least three. 
 In recent years there were several proposals aiming to extend the set of control parameters in Rydberg atoms\cite{PhysRevX.12.011040,PhysRevLett.128.083202,2015NJPh...17l3017L,2016PhRvA..94e1603L,PhysRevB.108.125135,PhysRevB.108.075146,PhysRevB.106.155411}. Among them the multi-species\cite{PhysRevX.12.011040,PhysRevLett.128.083202} and the multi-component\cite{2015NJPh...17l3017L,2016PhRvA..94e1603L} Rydberg arrays seems the most promising. In multi-species setup  a certain arrangements of Cs and Rb atoms is prepared. Rabi frequency, laser detuning and van der Waals potential can be individually controlled for each individual  specimen in addition to the inter-species interaction, so the effective model spans over six-dimensional parameter space. In multi-component Rydberg atoms, only one type of atoms is used  but each atom can be excited to one of the two Rydberg levels - components as sketched in Fig.\ref{fig:sketch}. The levels can be selected such that (i)
first-component interaction is much stronger than the second component one; and (ii) all other interactions including those between different components are negligibly small. 
 This model has five independent parameters: inter-atomic spacing, 
 %(that for a given pair of Rydberg levels defines the strength of the interaction for both components), 
two Rabi frequencies $\Omega_1$ and $\Omega_2$ and two individually controlled laser detunings $\Delta_1$ and $\Delta_2$ (see Fig.\ref{fig:sketch}).  Between two set-ups - multi-species and multi-component arrays - the latter has one significant advantage in the context of quantum phase transition and chiral melting: it preserves translation symmetry.

{\bf The blockade model.} Because of the very fast increase of the van der Waals potential at short distances simultaneous occupation of atoms within certain blockade radius is essentially excluded. It allows us to approximate long-range interactions of two components with two types of Rydberg blockade: nearest-neighbor blockade for the weak  and next-nearest-neighbor one for the strong components. This effectively fixes two parameters - relative interaction strength of two species and an interatomic distance.
The resulting model is defined by the following microscopic Hamiltonian acting in a constrained Hilbert space:
% \begin{multline}
%   H_{\text{MC}}=\sum_{\alpha=1,2}\sum_{i} \left[-\Omega_\alpha(d^\dagger_{\alpha,i}+d_{\alpha,i})-\Delta_\alpha n_{\alpha,i}\right]\\
%   n_{1,i}n_{1,i+1}=n_{2,i}n_{2,i+1}=n_{2,i}n_{2,i+2}=n_{1,i}n_{2,i}=0.
%     \label{eq:hamilt2}
% \end{multline}
\begin{subequations}
\begin{equation}
  H_{\text{MC}}=\sum_{\alpha=1,2}\sum_{i} \left[-\Omega_\alpha(d^\dagger_{\alpha,i}+d_{\alpha,i})-\Delta_\alpha n_{\alpha,i}\right]
  \label{eq:hamilt2}
 \end{equation}  
  \begin{equation}
  n_{1,i}n_{1,i+1}=n_{2,i}n_{2,i+1}=n_{2,i}n_{2,i+2}=n_{1,i}n_{2,i}=0.
    \label{eq:constraint}
\end{equation}  
\end{subequations}
Here $d^\dagger_{\alpha,i}$ brings an atom at site $i$ from the ground-state to a first or second Rydberg level  if $\alpha=1$ or $2$ correspondingly. This model has three independent parameters that we define as $\Delta_1/\Omega_1$, $\Delta_2/\Omega_2$, and $\Omega_1/\Omega_2$. We study the model with state-of-the-art DMRG algorithm\cite{dmrg1,dmrg2,dmrg3,dmrg4} with up to $N=907$ sites keeping  up to 2500 states and performing up to 8 sweeps (see Supplemental materials\cite{SM} for details).

\begin{figure}[t!]
\centering 
\includegraphics[width=0.45\textwidth]{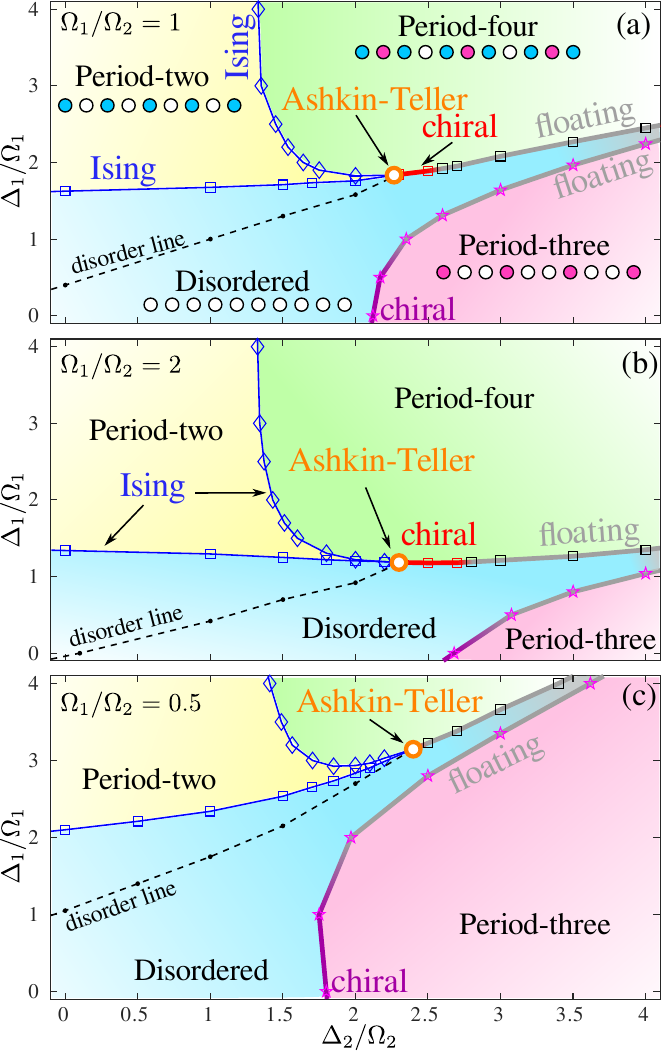}
\caption{Phase diagrams of the blockade models defined by Eqs.\ref{eq:hamilt2} and \ref{eq:constraint} as a function of laser detuning of two species and for three relative ratio of Rabi frequencies $\Omega_1/\Omega_2$. Each phase diagram contains four gapped phases: disordered phase, and three density wave-phases with periodicity $p=2,3$ and $4$. Typical patterns of these phases are sketched in (a) where white circles denote atoms in the ground state, blue and red circles state for atoms excited to the first (weak) and second (strong) Rydberg levels correspondingly. The period-2 phase is separated from the disordered and period-4 phases by two Ising transitions (blue squares and diamonds). The multi-critical point (open orange circle) belongs to the Ashkin-Teller universality class. For some interval in (a) and (b) the transition to the period-4 phase is direct and chiral (red line). The key observation is that the extent of the chiral transition in (b) for $\Omega_1/\Omega_2=2$ is significantly larger than in (a) for  $\Omega_1/\Omega_2=1$,  while for $\Omega_1/\Omega_2=0.5$ in (c)  the floating phase (gray) opens immediately. 
 The transition between period-three phase and the disordered phase is chiral  (purple line) for small values of $\Delta_1/\Omega_1$ and through a floating phase (gray) for large detuning of the weak component (see "Methods " for details).  }
\label{fig:pd}
\end{figure}

{\bf Phase diagram.} Our main results are  summarized in three phase diagrams in Fig.\ref{fig:pd}.
When laser detunings are small $\Delta_1/\Omega_1, \Delta_2/\Omega_2\ll 1$ the system is in the disordered phase and translation symmetry is not broken. For $\Delta_2/\Omega_2\gg\Delta_1/\Omega_1$ the system is populated with the strong component that due to  next-nearest-neighbor Rydberg blockade leads to a period-three phase separated from the disorder phase by either chiral transition or the floating phase\cite{SM}. In the opposite limit $\Delta_2/\Omega_2\ll\Delta_1/\Omega_1$ the system is in the period-two phase with every other site occupied by a weak component; the transition to the disordered phase is in the Ising universality phase\cite{SM}.
Upon increasing the detuning $\Delta_2$ the system undergoes the second Ising transition\cite{SM} where the translation symmetry is spontaneously broken once again and every other empty site of the period-2 phase is occupied by the strong component resulting in the period-four phase (sketched in Fig.\ref{fig:pd}(a)) with broken $\mathbb{Z}_2\times \mathbb{Z}_2$ symmetry (see Supplemental materials\cite{SM} for further details). Upon increasing $\Delta_2/\Omega_2$ two Ising lines come closer and eventually merge into a multicritical point  in the Ashkin-Teller universality clas\cite{difrancesco}. Beyond this point the transition from the disordered to the period-four phases is either direct in the chiral universality class or via the floating phase as shown in Fig.\ref{fig:pd}. This is further suported by incommensurate correlations that weak component develops beyond the disorder line\cite{SM}.

We distinguish three types of transitions - Ashkin-Teller, chiral and floating phase -  by looking at the product $|q-\pi/2|\times\xi$, where $q$ is incommensurate wave-vector  approaching its commensurate value with the critical exponent $\bar{\beta}$ and $\xi$ is a correlation length diverging with the critical exponent $\nu$.
To the best of our knowledge for the Ashkin-Teller criticality the exact value of $\bar{\beta}$ is not known but according to Huse and Fisher $\bar{\beta}>\nu$\cite{HuseFisher,HuseFisher1984}. Then for the Ashkin-Teller point the product $|q-\pi/2|\times\xi$ is expected to vanish. 
By contrast, at the chiral transition the equality $\nu=\bar{\beta}$  should hold and $|q-\pi/2|\times\xi$ takes some finite value. In case of the floating phase, it is separated from the disordered phase by the Kosterlitz-Thouless transition\cite{Kosterlitz_Thouless} characterized by the stretch-exponential divergence of correlation length, at the same time the wave-vector $q$ remains incommensurate across the transition, therefore $|q-\pi/2|\times\xi$ diverges.

\begin{figure*}[t!]
\centering 
\includegraphics[width=0.99\textwidth]{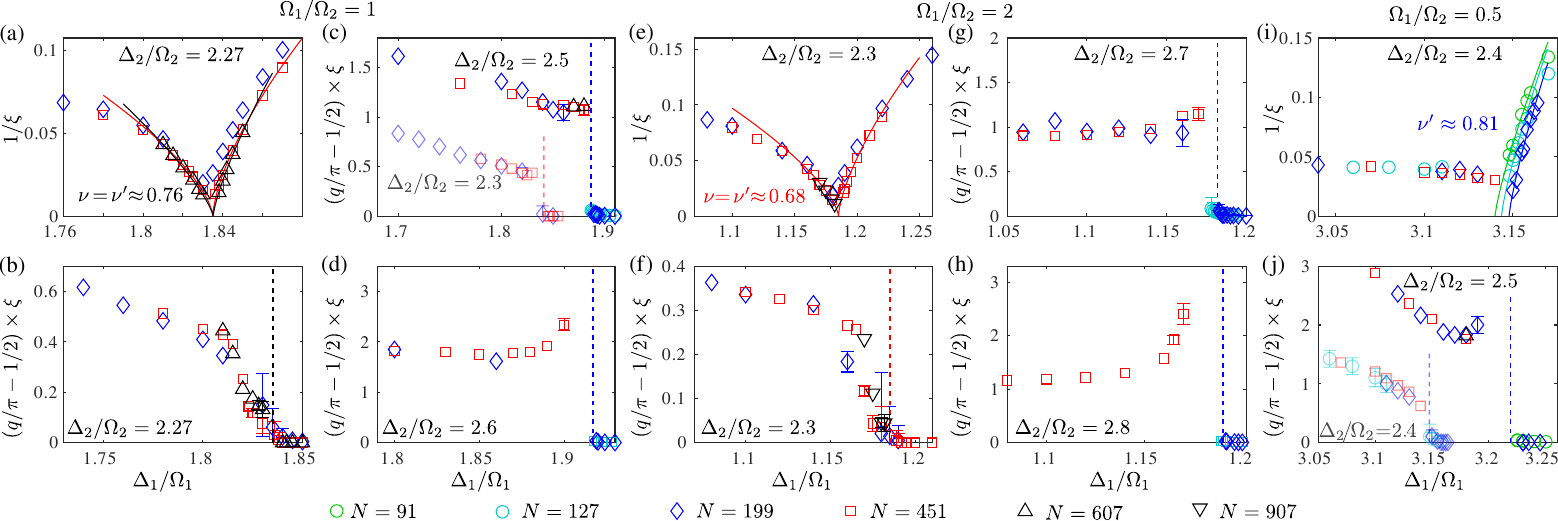}
\caption{ Inverse correlation length $1/\xi$ and product $|q-\pi/2|\times\xi$ along various cuts across the transition {\bf (a)-(d)} for $\Omega_1/\Omega_2=1$; {\bf (e)-(h)} for $\Omega_1/\Omega_2=2$; and {\bf (i)-(j)} for $\Omega_1/\Omega_2=0.5$.  {\bf (a),(b),(e),(f),(i)} and pale symbols in {\bf(j)}: vertical cut through the Ashkin-Teller point. {\bf (c),(g)}: vertical cuts  through chiral transitions. {\bf (d),(h)} and bright symbols in {\bf (j)}: a cut through a floating phase. In {\bf (a)} and {\bf (e)} correlation lengths are fitted with the power law with equal critical exponents $\nu=\nu^\prime$ specified in each panel.  $|q-\pi/2|\times\xi$ is defined with error bars $2\times \xi^2/N^2$; we only show errorbars if they exceed the size of the symbols. Dashed lines show the boundary of the ordered phase extracted by fitting the correlation length, color code corresponds to the legend in the lower part of the figure.}
\label{fig:cor1}
\end{figure*}

In Fig.\ref{fig:cor1} we show the inverse of the correlation length and the product $|q-\pi/2|\times\xi$ across different cuts through the transition. For $\Omega_1/\Omega_2=1$ and $\Delta_2/\Omega_2=2.27$ presented in  Fig.\ref{fig:cor1}(a),(b) the divergence of the correlation length is symmetric and numerically extracted critical exponent $\nu=\nu^\prime\approx 0.76\pm0.02$ is consistent with the Ashkin-Teller critical theory; the product $|q-\pi/2|\times\xi$ vanishes.
  This Ashkin-Teller point belongs to the interval where chiral perturbations results into a chiral transition\cite{nyckees2022commensurate,luescher}. Indeed, already at $\Delta_2/\Omega_2=2.3$ the product $|q-\pi/2|\times\xi$ shown in Fig.\ref{fig:cor1}(c) takes some finite value at the transition. This picture remains valid up to $\Delta_2/\Omega_2=2.5$. The location of the transition is extracted by fitting the correlation length in the period-four phase\cite{SM}. For $\Delta_2/\Omega_2=2.6$ presented in Fig.\ref{fig:cor1}(d) we see  a sign of divergence of $|q-\pi/2|\times\xi$ signaling the opening of the floating phase.

% \begin{widetext}

% \end{widetext}
%

For the larger ratio of Rabi frequencies  $\Omega_1/\Omega_2=2$ the location of the  Ashkin-Teller point is almost the same but the extracted critical exponent is noticeably smaller $\nu\approx0.68\pm0.04$ (see Fig.\ref{fig:cor1}(e),(f)). Beyond this point we detect the chiral transition that extends at least up to $\Delta_2/\Omega_2=2.7$ as shown in Fig.\ref{fig:cor1}(g). In Fig.\ref{fig:cor1}(h) for $\Delta_2/\Omega_2=2.8$ one can clearly see a divergence of $|q-\pi/2|\times\xi$ upon approaching the floating phase.  The larger interval of the chiral transition compare to the previous case with $\Omega_1/\Omega_2=1$  is fully consistent with smaller critical exponent $\nu$ at the Ashkin-Teller point(see Fig.\ref{fig:cor1}(e)). In other words, by increasing  $\Omega_1/\Omega_2$ one can tune the multicritical Ashkin-Teller point towards larger $\lambda$ and smaller $\nu$, that in turns lead to a longer chiral transition as sketched in Fig.\ref{fig:sketch3d}. Note also that the difference between the chiral transition and floating phase is more pronounced for $\Omega_1/\Omega_2=2$.

For  $\Omega_1/\Omega_2=0.5$ the multicritical point  is very close to the period-three phase and the floating phase surrounding it, as a consequence the correlation length is very large all over the narrow window of the disordered phase. This prevents us from fitting the divergence of the correlation length inside the disordered phase but the product $|q-\pi/2|\times\xi$ computed locally remains a valid measure: it goes to zero at the Ashkin-Teller point at $\Delta_2/\Omega_2=2.4$ (see Fig.\ref{fig:cor1}(i)-(j)), while for  $\Delta_2/\Omega_2=2.5$ it already shows a signature of divergence suggesting the presence of the floating phase. The latter is further supported by the divergence of the correlation length with the Pokrovsky-Talapov critical exponent $1/2$ (see Supplemental materials \cite{SM}). The picture is completed by the observation that along the cut  $\Delta_2/\Omega_2=2.4$ through the Ashkin-Teller point the correlation length diverges with the critical exponent $\nu^\prime\approx0.81$ that lies outside of the interval $2/3\leq \nu\lesssim 0.8$ where chiral transition is possible. Of course, neither this interval nor the critical exponents $\nu^\prime$ are exact and we cannot exclude a possibility of a short chiral transition between  $\Delta_2/\Omega_2=2.4$ and $2.5$. But what we observe here is a very important tendency of the Ashkin-Teller critical exponent to increase with decreasing $\Omega_1/\Omega_2$ such that the chiral transition shrinks to the point when it eventually disappears.

{\bf Discussion.} To summarize, our results offer a novel approach to manipulate quantum criticality in Rydberg atoms that, in particular, overcomes the bottleneck associated with the short extent of the chiral transition.
In addition, we formulate a protocol to probe the complete phase diagram of the chiral Ashkin-Teller model. 
Altogether, these provides an opportunity to explore universal and non-universal properties of chiral transitions - a necessary step towards a formal definition of the $p=4$ chiral universality class. 
 We expect our results obtained for the blockade approximation to be qualitatively correct for the model with  $1/r^6$ potential as soon as the selected Rydberg levels and the lattice spacing are such that detuning of weak and strong components alone leads to stable period-two and period-three phases correspondingly. 
 Multi-component Rydberg arrays opens a unique opportunity to explore fusion rules of quantum criticalities directly on a lattice. Here we focused on the simplest case: fusion of two Ising transitions into the Ashkin-Teller point followed by the chiral transition. Playing with different interaction regimes and longer blockades one can realize, for instance, a fusion of chiral transitions. 
%   It would be interesting to explore models with different interaction regimes, in particular, multi-component Rydberg arrays with longer blockade radii that offers a unique opportunity to explore fusion rules of chiral transitions.
Further generalization to three and more components is conceptually straightforward and introduces a new class of models  with multi-component Hilbert space. Multiple individually controllable parameters enables a  fine-tuning, for instance, to points with higher symmetries.

 \section{Acknowledgments}

I thank Bernhard L\"uscher, Samuel Nyckees and Frederic Mila for insightful discussions on the chiral Ashkin-Teller model and to Jiri Minar for inspiring discussion on multi-component Rydberg arrays.
This work has been supported by the Delft Technology Fellowship and it was performed in part at Aspen Center for Physics, which is supported by National Science Foundation grant PHY-2210452.
 Numerical simulations have been performed at the DelftBlue HPC and at the Dutch national e-infrastructure with the support of the SURF Cooperative.

\bibliographystyle{apsrev4-1}
\bibliography{bibliography,comments}

\newpage

 {Supplemental matherial to "Tunable quantum criticality in multi-component Rydberg arrays"}

\section{Details about the algorithm}

\begin{figure}[h]
\centering 
\includegraphics[width=0.5\textwidth]{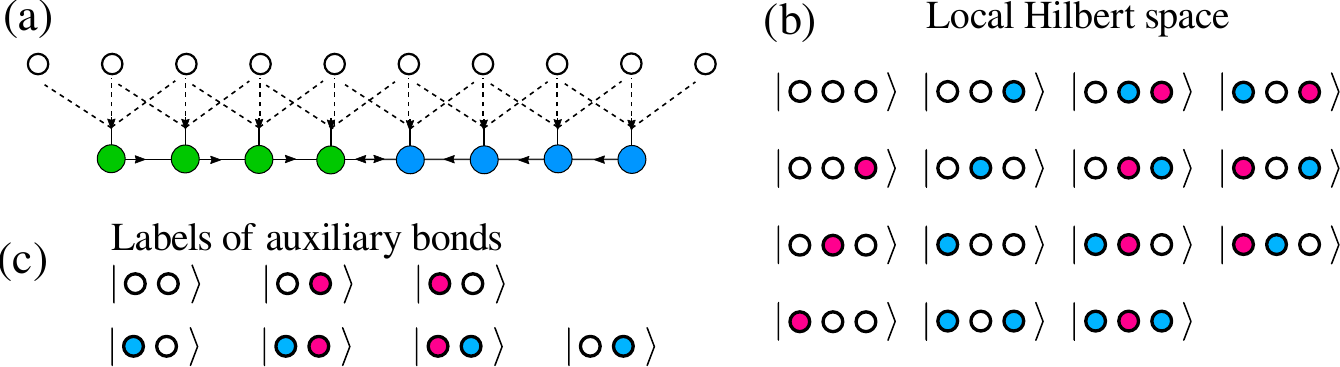}
\caption{ {\bf Implementation of DMRG with two-site blockade.} (a) We associate MPS tensor with three consequtive Rydberg atoms. The Hilbert space of this three-site object has dimension $d=17$ sketched in (b). Every pair of MPS share two Rydberg atoms, the states of these two atoms sketched in (c) are used to label auxiliary legs of MPS. }
\label{fig:dmrgblock}
\end{figure}

We perform numerical simulations with DMRG algorithm in the form of variational optimization of the matrix product states (MPS)\cite{dmrg1,dmrg2,dmrg3,dmrg4}. We implement the blockade by applying a strong penalty terms $U(n_{1,i}n_{1,i+1}+n_{2,i}n_{2,i+1}+n_{2,i}n_{2,i+2})$. We choose the penalty to be $U=400$: this value is sufficiently strong to keep expectation value of any of the three terms listed above below $10^{-6}$ even in the vicinity of the transition, while this value is not too large to cause instabilities and convergence issues. We also benchmarked our algorithm with constrained DMRG with explicitly implemented blockades\cite{chepiga2021kibble,scipost_chepiga}: the two sets of results do not have any noticeable differences. Quite surprisingly the unconstrained DMRG with the penalty terms turns out to be more efficient for numerical simulations of multi-component Rydberg arrays than the constrained version of the algorithm that by construction obeys a block-diagonal structure. 
In order to implement constrained DMRG for multi-component system, we generalize the implementation of next-nearest-neighbor blockade described in details Ref.\cite{chepiga2021kibble}. We span local degrees of freedom over three consecutive Rydberg atoms as sketched in Fig.\ref{fig:dmrgblock}(a). This increase the local Hilbert space to $d=17$ listed in Fig.\ref{fig:dmrgblock}(b), note that out of $3^3=27$ total states 10 states violating the blockades were excluded. In this construction, every pair of consequtive MPS have share two Rydberg atoms, therefore one can use the state of these atoms as quantum labels that create one-to-one correspondence between auxiliary legs of MPS. Thes seven labels are listed in Fig.\ref{fig:dmrgblock}(c).

Contracting the networks block-by-block allows to reduce the memory cost significantly, however an auxiliary bond of tensors having seven different labels leads to a large number of rather small blocks such that an over-head for a block-by-block operation turns out to be higher than the gain from the lower memory cost.  In DMRG simulations we reach the convergence of the systems with up to $N=907$ sites keeping  up to 2500 states and performing up to 8 sweeps; we generically discard singular values below $10^{-8}$. 

\section{Additional details on various phases and boundary conditions}

\begin{figure}[h!]
\centering 
\includegraphics[width=0.5\textwidth]{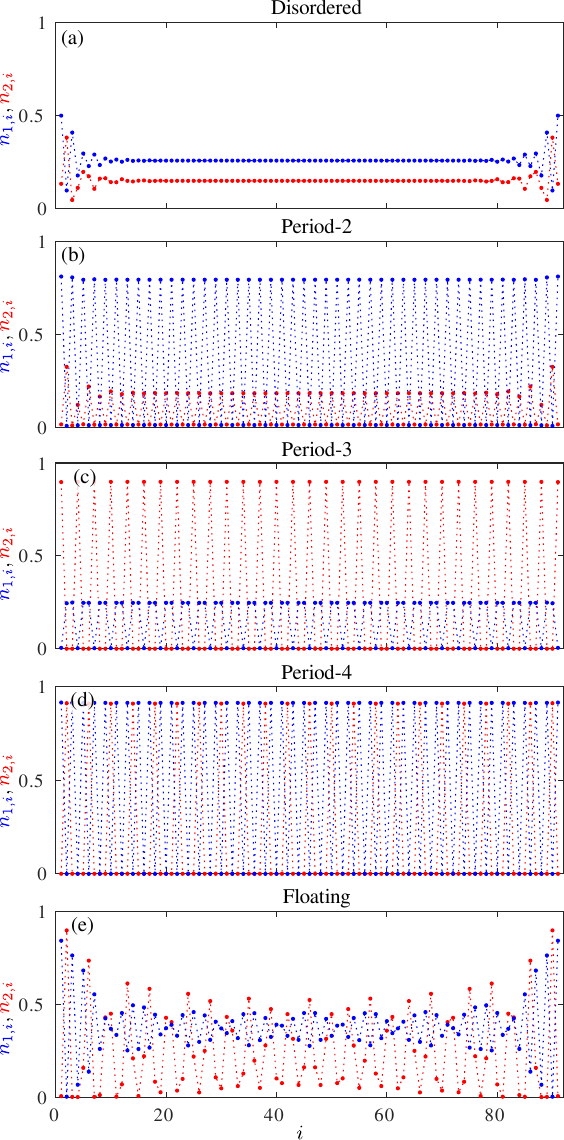}
\caption{ {\bf Local density of the strong(red) and weak (blue) component in various phases}. Density of (a)  the disordered phase at $\Delta_1/\Omega_1=1$ and $\Delta_2/\Omega_2=1$; (b) the period-2 phase at  $\Delta_1/\Omega_1=2$ and $\Delta_2/\Omega_2=0$; (c) the period-3 phase at  $\Delta_1/\Omega_1=0$ and $\Delta_2/\Omega_2=3$; (d) the period-4 phase at $\Delta_1/\Omega_1=3$ and $\Delta_2/\Omega_2=3$; (e) the floating phase in the vicinity of the period-4 phase at $\Delta_1/\Omega_1=2$ and $\Delta_2/\Omega_2=3$}
\label{fig:phases}
\end{figure}

In this section we provide further details on various phases and on the used boundary conditions.
As explained in the main text, there are four gapped phases: disordered, and period-2, 3 and 4. 
% In addition there are two floating phases that evenuallu can merge into one for large $\Delta_2/\Omega_2$.
Fig.\ref{fig:phases} present local density profiles along a chain with $N=91$ sites at various characteristic points of the phase diagram for $\Omega_1/\Omega_2=1$. Fig.\ref{fig:phases}(a) shows the density profile at  $\Delta_1/\Omega_1=1$ and $\Delta_2/\Omega_2=1$ in the disordered phase. One can clearly see that in the bulk the density of each component is uniform. Note that the microscopic model does not preserve the number of particles in the weak/strong component, so the density of each component changes continuously all over the disordered phase.  

Fig.\ref{fig:phases}(b) presents the density profile at  $\Delta_1/\Omega_1=2$ and $\Delta_2/\Omega_2=0$ in the period-2 phase where every other site is occupied by a weak (blue) component. In the thermodynamic limit this phase is two-fold degenerate - weak component can occupy either even or odd sites. In addition one can still detect a small density of the strong (red) component uniformly distributed among sites that are not occupied by a weak component.

Fig.\ref{fig:phases}(c) shows the period-3 phase at   $\Delta_1/\Omega_1=0$ and $\Delta_2/\Omega_2=3$. Every third site is occupied by the strong (red) component, while the remaining sites host a certain density of the weak component uniformly distributed over unoccupied sites. This phase breaks $\mathbb{Z}_3$ symmetry and in the thermodynamic limit it has three-fold degenerate ground-state. 

The density of the period-4 phase taken at  $\Delta_1/\Omega_1=3$ and $\Delta_2/\Omega_2=3$ is shown in Fig.\ref{fig:phases}(d). One can see that every second site is occupied by the weak (blue) component, while every other of the remaining sites are occupied by the strong (red) component. This phase spontaneously breaks $\mathbb{Z}_2\times \mathbb{Z}_2$ symmetry and has a four-fold degenerate ground-state in the thermodynamic limit.

Finally, in Fig.\ref{fig:phases}(e) we also present a typical density profile of the floating phase taken in the vicinity of the period-4 phase at  $\Delta_1/\Omega_1=2$ and $\Delta_2/\Omega_2=3$. Note that in the thermodynamic limit the density in this critical phase is uniform and the presence of incommensurability can only be detected via correlation functions. Oscillations in local density that we observe in Fig.\ref{fig:phases}(e) are visible because of explicitly broken translation symmetry due to open boundary conditions.

As one can see in  Fig.\ref{fig:phases}(b)-(d), ordered phases favor occupied sites at the edges. This impose boundary conditions to be respected in numerical simulations: for the period-2 phase the total number of sites should satisfy $N=2k+1$ with integer $k$. Period-3 phase requires $N=3k+1$. In the period-4 phase the edges favor weak-strong-weak cluster over the first and last three sites implying that the chain length should satisfy $N=4k+3$. As one can easily check $N=91,199,451,607,907$ used in this paper satisfy all three conditions.

\section{Location of the Ising transitions}

\begin{figure}[h]
\centering 
\includegraphics[width=0.5\textwidth]{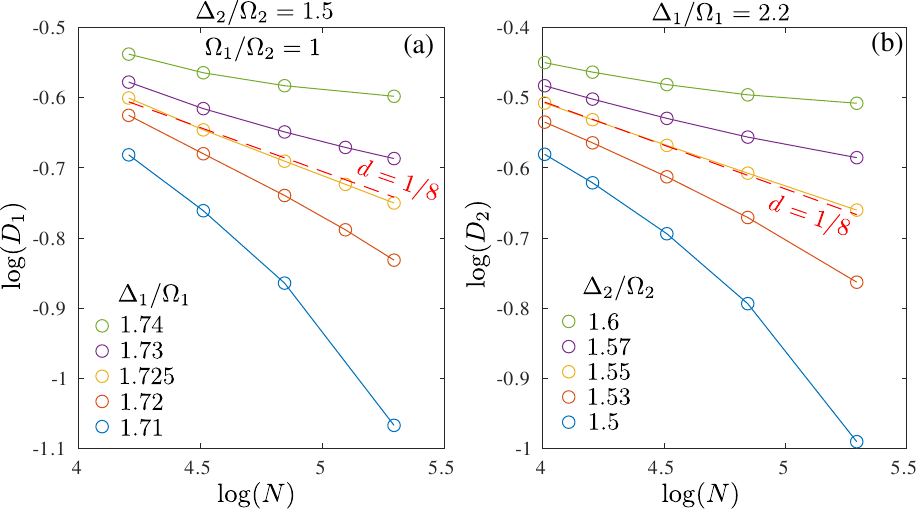}
\caption{ {\bf Finite-size scaling of the local density oscillations across two Ising transitions.}  Scaling of the local density oscillations in (a) weak component across the transition between the disordered and period-two phases and (b) in strong component across the transition between the period-two and period-four phases. In each case the separatrix is associated with the transition and the slope agree with the Ising scaling dimension $d=1/8$ (dashed red line) shown for a reference. }
\label{fig:ising}
\end{figure}

In order to locate Ising transitions we look at the finite-size scaling of the density oscillations - a local measure of spontaneously broken translation symmetry. In the period-two phase the translation symmetry is broken by the weak component and by two sites, so the relevant operator is defined as $D_1=|\langle n_{1,j}\rangle-\langle n_{1,j+1}\rangle|$ that we compute in the middle of the chain. We look at this quantity in the log-scale and associate the transition with the separatrix that separates disordered phase (concave curves) and the symmetry broken phase (convex ones) as shown in Fig.\ref{fig:ising}(a). The slope of the separatrix corresponds to the scaling dimension that is in excellent agreement with the conformal field theory prediction for the Ising transition $d=1/8$\cite{difrancesco}.

At the transition to the period-four phase the translation is again spontaneously broken: every other site that was not occupied by the weak component in the period-two phase becomes occupied by the strong component in the period-four phase. So the local order parameter can be defined as $D_2=|\langle n_{2,j}\rangle-\langle n_{2,j+2}\rangle|$, where $j,j+2$ are sites that are not occupied by the weak component. As previously, we associate the transition with the separatrix and its slope is in excellent agreement with the Ising scaling dimension $1/8$ (see Fig.\ref{fig:ising}(b)).

We cross-check the nature of these two transitions by extracting the central charge.   The entanglement entropy $S_N(n)=-\mathrm{Tr} \rho_n \ln \rho_n$ is extracted from the eigenvalues of the reduced density matrix $\rho_n$.
We extract central charge numerically from the finite-size scaling of the entanglement entropy in an open chain that on a finite chain with $N$ sites scales with the size of the subsystem $n$ ($1\ll n\ll N$) as\cite{CalabreseCardy}:
\begin{equation}
{S}_N(n)=\frac{c}{6}\ln d(n)+s_\mathrm{B},
\label{eq:calabrese_cardy}
\end{equation}
where $d=\frac{2N}{\pi}\sin\left(\frac{\pi n}{N}\right)$ is the conformal distance and  $s_\mathrm{B}$ is a non-universal term that includes, in particular, the boundary entanglement entropy. The scaling of the entanglement entropy for the two critical points identified in Fig.\ref{fig:ising}(a),(b) are presented in Fig.\ref{fig:ising}(a) and (b) correspondingly. In both cases the numerically extracted values are in a good agreement with the theory predictions $c=1/2$. Small discrepancy is due to a finite resolution of the location of the critical point.

\begin{figure}[h]
\centering 
\includegraphics[width=0.5\textwidth]{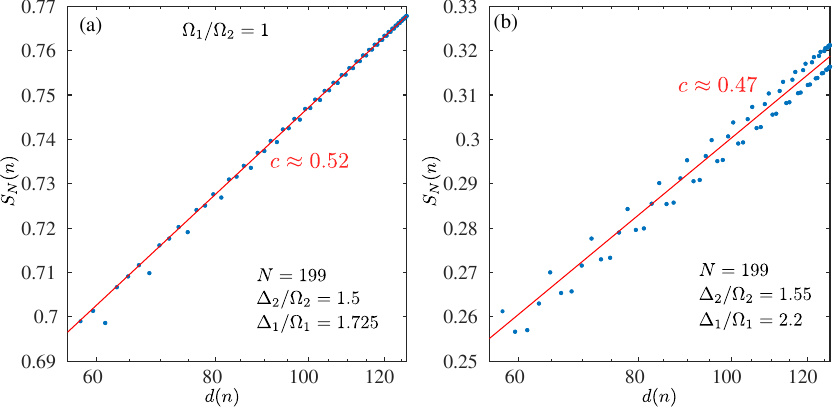}
\caption{ {\bf Scaling of entanglement entropy at the two Ising transitions.} In each case in each case the central charge extracted with Calabrese-Cardy formula in Eq.\ref{eq:calabrese_cardy} agrees with theory prediction $c=1/2$ for the Ising transition.}
\label{fig:dis}
\end{figure}

\section{Extraction of the correlation length and of the wave-vector}
\begin{figure}[h]
\centering 
\includegraphics[width=0.5\textwidth]{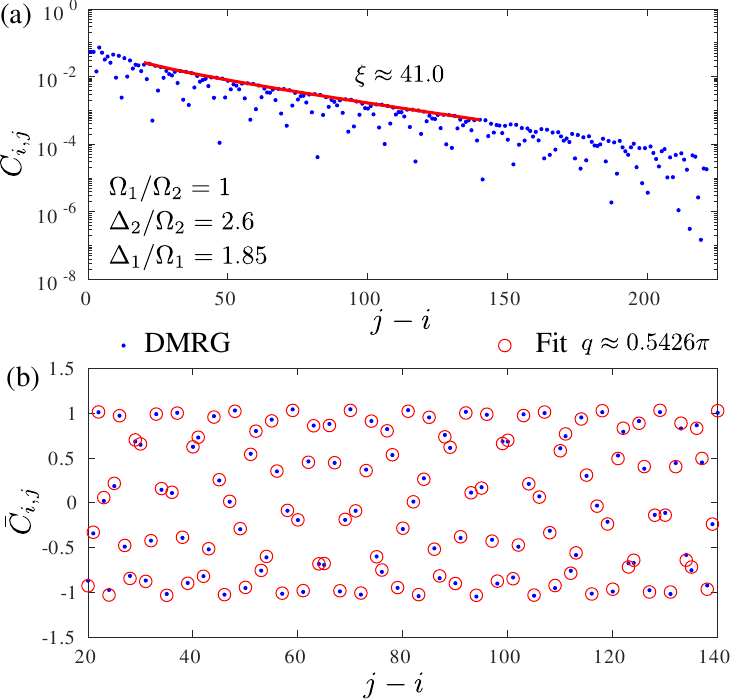}
\caption{ {\bf Example of  fit of the correlation function to the Ornstein-Zernicke form.} In the first step (a), we extract the correlation length discarding the oscillations. In the second step (b), we fit the reduced correlation function and extract the wave-vector $q$.}
\label{fig:IncomFit}
\end{figure}

In order to extract the correlation length and the wave-vector $q$, we fit the connected correlation function for the strong component $C_{i,j}=\langle n_{2,i}n_{2,j}\rangle-\langle n_{2,i}\rangle\langle n_{2,j}\rangle$ to the Ornstein-Zernicke form\cite{ornstein_zernike}:

\begin{equation}
C^\mathrm{OZ}_{{i,j}}\propto \frac{e^{-|i-j|/\xi}}{\sqrt{|i-j|}}\cos(q|i-j|+\varphi_0),
\end{equation}
where the correlation length $\xi$, the wave vector $q$, and the initial phase $\varphi_0$ are fitting parameters. We extract the correlation length and the wave-vector in two steps.  First, we discard the oscillations and fit the main slope of the decay as shown in Fig.\ref{fig:IncomFit}(a) such that the fit can be performed in a semi-log scale $\log C(x=|i-j|)\approx c-x/\xi-\log(x)/2$. This provides more accurate estimates of the correlation length on a long scale. In fact, there are two very similar strategies to perform this fit: either to fit all the points assuming that oscillations are fast enough to cancel; or to fit only top of the arcs assuming that there are enough arcs  for a reliable fit. Here we use the second method and include the point into a fit if it exceeds two points on its left and two points on its right. Let me emphasize here that both methods are equivalent and affect only an amplitude and a centering of a cosine function in the next step.

In the second step we fit the reduced correlation function
 \begin{equation}
\tilde{C}_{i,j}=C_{i,j} \frac{\sqrt{|i-j|}}{e^{-|i-j|/\xi+c}}
\end{equation}
with a cosine $\tilde{C}_{i,j}\approx a\cos(q|i-j|+\varphi_0) $ as shown in Fig.\ref{fig:IncomFit}(b). The agreement between the DMRG data (blue dots) and the fit (red circles) is almost perfect.
% By fitting the correlation over different intervals we find that an error in correlation length do not exceed $3\%$, and the wave-vector $q$ is determined with the precision O$(10^{-6})$.
 We estimate error in $q$ caused by finite-size effects as $2\pi\xi/N^2$, where $2\pi/N$ is a natural step in $q$ due to fixed boundary conditions and assuming for infinite correlation length, an additional factor $\xi/N$ reflect the possibility to adjust $q$ at the edges when correlation length is finite.

\section{Transition out of period-three phase}

\begin{figure}[h]
\centering 
\includegraphics[width=0.5\textwidth]{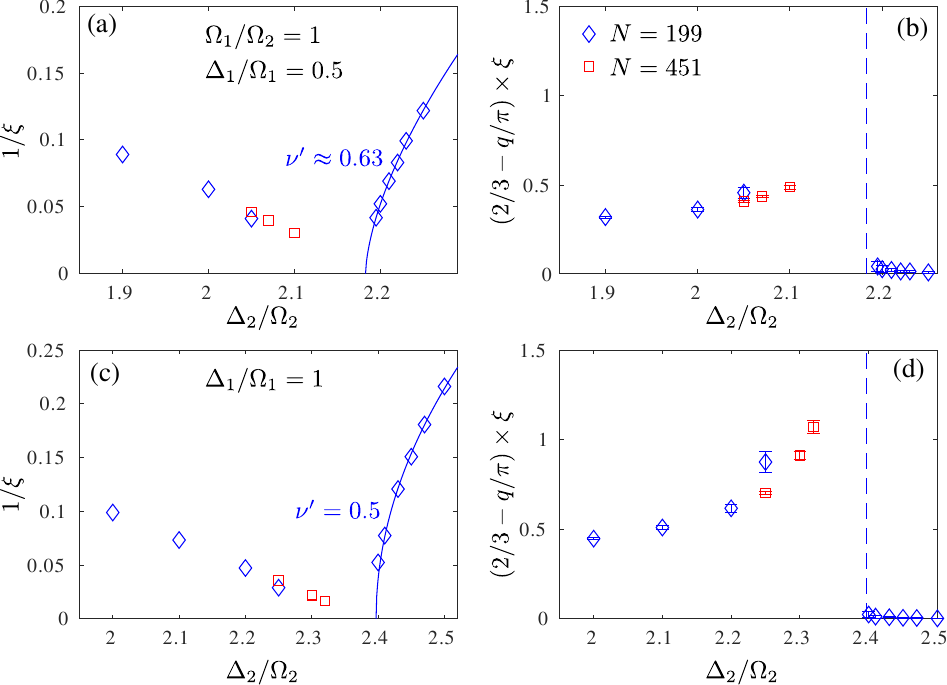}
\caption{ {\bf Inverse of the correlation length and the product $|2\pi/3-q|\times\xi$ along two horizontal cuts across the transition to the period-three phase for  $\Omega_1/\Omega_2=1$.} (a),(b) - a cut through a chiral transition point at $\Delta_1/\Omega_1=0.5$; (c),(d) a cut through the floating phase at $\Delta_1/\Omega_1=1$. In (a) and (c) the correlation lengths in the period-three phase are fitted with the power law.  Dashed blue lines show the boundary of the ordered phase extracted in (a) and (c).}
\label{fig:p3}
\end{figure}

In order to locate the boundary of the period-three phase and to identify the nature of the transition we follow a similar procedure as for the boundary of the period-four phase. We look at the product $|2\pi/3-q|\times\xi$ that is expected to take a finite value if transition is chiral and diverge in the case of intermediate floating phase. Here we expect conformal Potts point only in the limit $\Delta_1/\Omega_1\rightarrow\infty$. We also look at the scaling of the correlation length in the period-three phase. In Fig.\ref{fig:p3} we provide an example of two typical cuts through the transition for $\Omega_1/\Omega_2=1$: when it is chiral in Huse-Fisher universality class in Fig.\ref{fig:p3}(a),(b); and when it goes through an intermediate floating phase in Fig.\ref{fig:p3}(c),(d). Small increase of $|2\pi/3-q|\times\xi$ in Fig.\ref{fig:p3}(b) indicates its proximity to the Lifshitz point, where the floating phase opens.

Following the same procedure we identified the boundaries of the period-three phase for $\Omega_1/\Omega_2=2$ and $0.5$. For $\Omega_1/\Omega_2=2$ our results at $\Delta_1/\Omega_1=0$ are consistent with chiral transition, while already at $\Delta_1/\Omega_1=0.5$ we see a presence of the floating phase. For  $\Omega_1/\Omega_2=0.5$ chiral transition is stable up to $\Delta_1/\Omega_1=1$ and the floating phase is detected at $\Delta_1/\Omega_1=2$.

\section{Disorder line}

Chiral perturbations usually manifest itself in incommensurate correlations. The line where real-space correlations switch from commensurate to incommensurate regimes is usually called in the literature the disorder line. In order to locate this line in the disordered phase we look at the connected correlations of the weak component. A few examples of commensurate and incommensurate scalings are shown in Fig.\ref{fig:dis}; we associate the disorder line with the point where incommensurability disappear.  Calculations were performed with $N=67$  sites deep inside the disordered phase and with $N=91$ in the vicinity of the multicritical point. 

\begin{figure}[h]
\centering 
\includegraphics[width=0.5\textwidth]{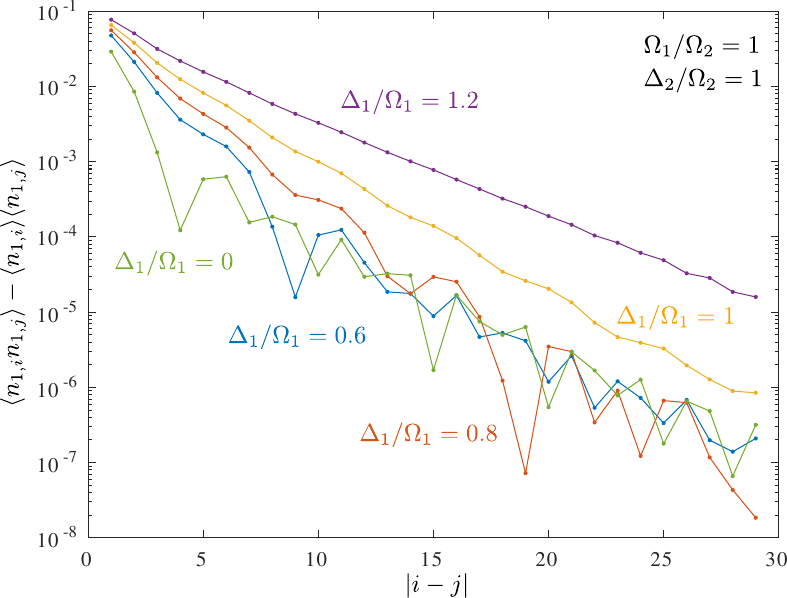}
\caption{ {\bf Location of the disorder line.} Scaling of the weak component correlations as a function of distance for $\Omega_1/\Omega_2=1$ and $\Delta_2/\Omega_2=1$. For $\Delta_1/\Omega_1<1$ one can clearly see the presence of incommensurability that disappears for $\Delta_1/\Omega_1>1$. }
\label{fig:dis}
\end{figure}

 It is not always clear whether the system is commensurate or incommensurate (see, for instance, the data for $\Delta_1/\Omega_1=1$  in Fig.\ref{fig:dis}) and for this reason there is an uncertainty in the location of disorder line.  However, our results are consistent with the disorder line hitting the period-four phase at multi-critical point. This agree with expectation that system is commensurate on both sides of the Ising transition.  
 In fact the situation is more subtle here. There are two components and two commensurate Ising transitions. At the transition between the disordered and the period-two phase the weak component breaks translation symmetry and this particular component has to be commensurate in the vicinity of the transition, however the strong component can remain incommensurate across this Ising transition. But at the second Ising transition from period-two to period four phases it is now the strong component that breaks the translation symmetry and thus it has to be commensurate in the vicinity of the transition. It implies that in principle there must be a second disorder line where incommensurability develops in the strong component; this line can be located in the period-two or in the disordered phase and also expected to end at the multicritical Ashkin-Teller point.

\section{Divergence of the correlation length at the transition to the period-four phase}

\begin{figure}[h]
\centering 
\includegraphics[width=0.5\textwidth]{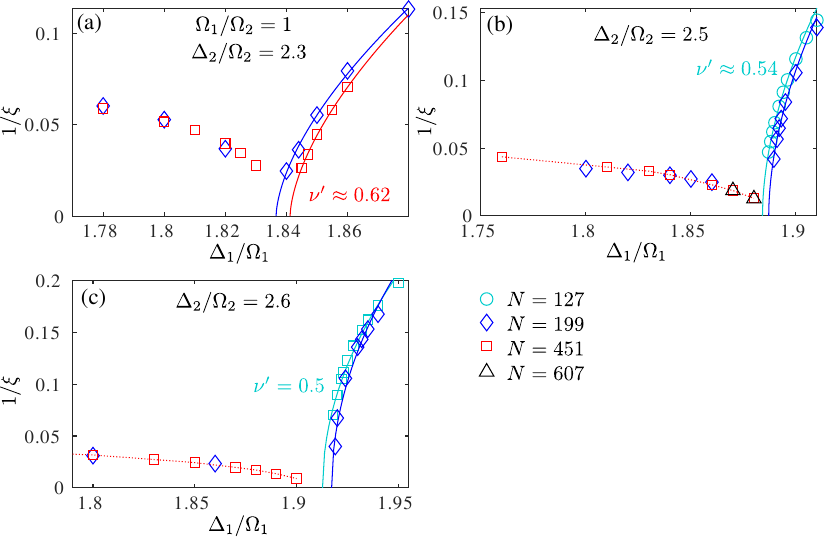}
\caption{ Inverse of the correlation length on both sides from (a,b) the chiral transition and from  (c) the floating phase for $\Omega_1/\Omega_2=1$. In each case the divergence of the correlation length in the period-four phase is fitted with a power law with the critical exponent $\nu^\prime$. }
\label{fig:coradd1}
\end{figure}

In the main text we distinguished three critical regimes - Ashkin-Teller, chiral and floating phase - by looking at the product $|q-\pi/2|\times\xi$. In addition we also extract the critical exponent $\nu^\prime$ that controls the divergence of the correlation length upon approaching the boundary of the period-four phase. At the Ashkin-Teller transition $2/3\leq \nu^\prime\leq 1$ and the corresponding results have been presented in Fig.4 (a),(e) and (i) in the main text. For the Pokrovsky-Talapov transition this critical exponent is expected to take the value $\nu^\prime=1/2$. For chiral transition the value of this exponent is not known. For $\Omega_1/\Omega_2=1$ and  for the two cuts that we identified as a chiral transition based on $|q-\pi/2|\times\xi$ the extracted critical exponents $\nu^\prime$ noticeably exceeds the value $\nu^\prime=1/2$ of the Pokrovsky-Talapov transition. The results are presented in Fig.\ref{fig:coradd1}(a),(b). By contrast, at $\Delta_2/\Omega_2=2.6$ that we identified as a cut through the floating phase the DMRG results are in excellent agreement with $\nu^\prime=1/2$.

 For $\Omega_1/\Omega_2=2$ and $\Delta_2/\Omega_2=2.7$ that we associated with chiral transition the critical exponent is about $\nu^\prime \approx  0.65$ (for smaller system sizes we see $\nu^\prime \approx 0.68$), while for $\Delta_2/\Omega_2=2.8$ we already see an excellent agreement with the expected square-root divergence. This further supports our conclusion that the Lifshitz point is located in the interval $2.7<\Delta_2/\Omega_2<2.8$.

\begin{figure}[h]
\centering 
\includegraphics[width=0.5\textwidth]{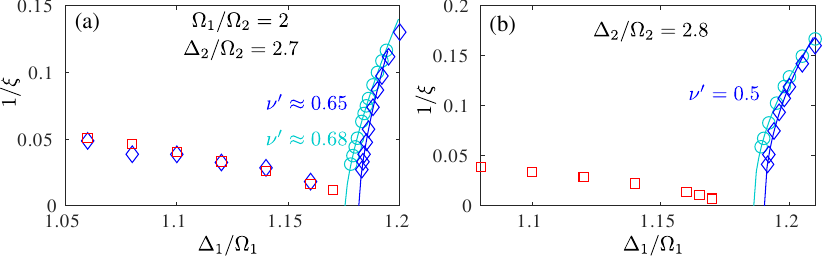}
\caption{ Inverse of the correlation length on both sides from (a) the chiral transition and (b) the floating phase for $\Omega_1/\Omega_2=2$ }
\label{fig:coradd2}
\end{figure}

\begin{figure}[h]
\centering 
\includegraphics[width=0.38\textwidth]{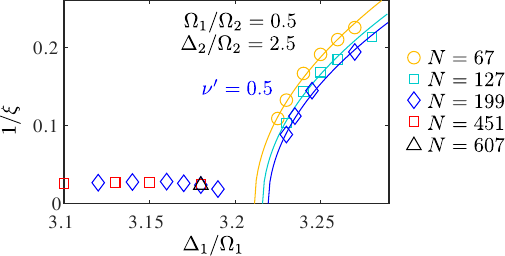}
\caption{ Inverse of the correlation length on both sides from (a) the chiral transition and (b) the floating phase for $\Omega_1/\Omega_2=0.5$ }
\label{fig:coradd05}
\end{figure}

 Finally, for $\Omega_1/\Omega_2=0.5$ already at $\Delta_2/\Omega_2=2.5$ we see that the correlation length diverges with the critical exponent $\nu^\prime=1/2$ as presented in Fig.\ref{fig:coradd05}.

 For all cuts we see a finite-size effect in the location of the transition. This effect is very common in Rydberg atoms\cite{PhysRevResearch.4.043102}. 

\end{document}